# High-Resolution X-Ray Studies of the Direct Spin Contact of EuO with Silicon

*Dmitry V. Averyanov[1], Andrey M. Tokmachev[1], Igor A. Likhachev[1], Eduard F. Lobanovich[1], Oleg E. Parfenov[1], Elkhan M. Pashaev[1], Yuri G. Sadofyev[1], Ilia A. Subbotin [1], Sergey N. Yakunin[1] & Vyacheslav G. Storchak[1*]*

[1]National Research Center "Kurchatov Institute", Kurchatov Square 1, Moscow 123182, Russia

ABSTRACT - Ferromagnetic semiconductor europium monoxide (EuO) is believed to be an effective spin injector when directly integrated with silicon. Injection through spin-selective ohmic contact requires superb structural quality of the interface EuO/Si. Recent breakthrough in manufacturing free-of-buffer-layer EuO/Si junctions calls for structural studies of the interface between the semiconductors. *Ex situ* high-resolution X-ray diffraction and reflectivity accompanied by *in situ* reflection high-energy electron diffraction reveal direct coupling at the interface. A combined analysis of XRD and XRR data provides a common structural model. The structural quality of the EuO/Si spin contact by far exceeds that of previous reports and thus makes a step forward to the ultimate goals of spintronics.

Spintronics, or spin electronics, is an emerging technology based on active control of spin degrees of freedom in solid-state systems.[1] Metal spintronics employing giant and tunnel magnetoresistance[2,3] presents a basis for numerous data storage and memory applications. Semiconductors have an advantage of being capable to amplify signals or transistor action thus leading to expectations of integrated memory and logic computing on the platform of semiconductor spintronics.[4] Rapidly increasing power consumption and heat generation of traditional electronic circuits based on transport of electric charge give new impetus to the search for alternative technologies. Semiconductor spintronics may just be a solution due to low energy dissipation when switching and computing.

So far, studies of GaAs dominated the field of semiconductor spintronics because of strong spin-orbit coupling and efficient optical orientation of spins in this material. However, silicon is a material of choice for modern electronics. Thus, it is not surprising that despite tremendous obstacles a huge concerted effort is aimed at developing silicon spintronics.[5] As in Si spin-orbit coupling is exceptionally weak and optical orientation is rather ineffective,[5] generation of sizable spin polarization (creation of a non-equilibrium spin population) in non-magnetic silicon is sought by *electrical* spin injection. However, direct injection of spin-polarized carriers from a ferromagnetic metal into Si is rather ineffective due to the impedance mismatch between the materials[6] albeit chemical interactions at the interface can also be of significance. A possible approach is to employ a tunable tunnel barrier,[7] which leads to spin polarization in Si even above room temperature.[8] Alternative ways of injection are based on hot electrons with energy well above the Fermi energy[9] and thermal spin flow due to Seebeck spin tunneling.[10] The spin polarization can be detected using silicon-based LED,[7] three-terminal (local)[8] and four-terminal (non-local) electrical devices.[11] However, despite all the efforts the combinations of spin polarization, spin current and spin decoherence time achieved to date are far from being sufficient for implementing spin functionality in silicon. When special attention is paid to the tunnel barrier some spin injection characteristics can be significantly improved[12,13] but the field still awaits a major breakthrough.

Electrical spin injection in a ferromagnetic (FM) semiconductor heterostructure[14] is an alternative not requiring a tunnel barrier and not suffering from the impedance mismatch problem. To explore this possibility one needs a stable epitaxial growth of a FM semiconductor directly on silicon. Dilute magnetic semiconductors make promising playground but their application is dogged by concerns regarding magnetic inhomogeneities[15], the possibilities of secondary FM phases, contamination issues, and the sensitivity of magnetism to growth conditions. Among semiconductors with inherent magnetic ordering, EuO is a leading candidate to be integrated with Si.[16-22] EuO has an advantage of being perfectly magnetically homogeneous in a FM state.[23] Huge exchange splitting of the conduction band (0.6 eV) makes EuO a dreamed-of source of almost fully spin-polarized electrons.[24,25] Band gap of EuO (1.12 eV) matches that of Si. Cubic EuO is structurally compatible with silicon. It is believed to be the only binary magnetic oxide thermodynamically stable in contact with Si.[26] Bulk EuO exhibits remarkable properties: colossal magnetoresistivity effect of about 6 orders of magnitude in a few Tesla field, metal-insulator transition accompanied by 13-15 orders of magnitude change in resistivity, pronounced magneto-optics effects, high sensitivity to strain[27] and doping. In short, EuO is a gem among magnetic materials.

In contrast, manufacturing EuO films directly on Si is notoriously difficult. The first challenge comes from a large lattice mismatch of 5.6% between EuO and Si. Growth of EuO from europium and oxygen is complicated by formation of higher oxides $Eu_3O_4$ and $Eu_2O_3$. Both Eu and $O_2$ interact with Si forming Eu silicide and silicon oxide, respectively. In addition, joining covalent and ionic systems can be very demanding. Therefore, numerous attempts to grow epitaxial EuO/Si heterostructures with a clean interface invariably fail, a typical outcome being formation of a polycrystalline EuO.[19,20] Impurity phases at the interface constitute another major problem.[21,22] Electron microscopy in combination with electron energy loss spectroscopy shows that EuO is separated from Si by a disordered layer of several nm with exceptionally high concentrations of parasitic phases including unwanted Eu silicide and those with non-magnetic $Eu^{3+}$.[22] EuO films of a better quality are produced when a buffer layer of SrO separates EuO and Si.[16,18] This, however, does not approach the ultimate goal of manufacturing EuO/Si spin-contact interface because an intermediate layer exponentially reduces the probability of spin-polarized carriers injection.

Recently, we proposed a novel approach to formation of a direct (free-of-buffer-layer) EuO/Si junction by application of a sub-monolayer of surface europium (or strontium) superstructure with (1×5) reconstruction.[28] This surface structure has a higher coverage of Si by metal atoms than used previously.[16,18] It becomes oxidized and incorporated into the oxide system at the very beginning of the growth.[29] In addition, we adopted a two-step protocol of EuO growth: the low-temperature phase comprising about 10 monolayers (MLs) of EuO diminishes formation of unwanted phases at the interface while the regular growth of EuO bulk is carried out in the high-temperature adsorption-controlled regime.[28] Initial magnetic and structural characterizations of EuO/Si films[28] indicate that this approach is capable of solving the long-standing problem of manufacturing direct EuO/Si spin contact interface.

The complexity of the system calls for thorough high-resolution structural studies. The large lattice mismatch between EuO and the substrate inevitably leads to significant distortions of the film. Their relaxation may affect the quality of the interface. The two-step growth procedure introduces additional irregularities into the EuO film.

What is most important, the structure of the interface determines its magnetic properties: for instance, the Curie temperature is highly susceptible to strains since exchange interaction is a steep function of Eu-Eu distance. Structural defects may produce magnetic inhomogeneities which may be highly destructive to spin injection.

In this Letter we present a detailed analysis of the structural coupling across the EuO/Si interface. A combination of *in situ* reflection high-energy electron diffraction (RHEED) and *ex situ* high-resolution X-ray diffraction (XRD) and reflectivity (XRR) studies accompanied by rigorous modeling reveals the structure of EuO films epitaxially integrated with silicon.

We discuss X-ray structural studies of two samples with similarly grown EuO/Si interfaces but different protective capping of the EuO thin film. Sample A is covered with amorphous SiO, while the topmost layer of Sample B is formed by higher oxide $Eu_2O_3$ manufactured by a controlled oxidation of EuO.

XRR and XRD studies of EuO/Si samples are carried out using Rigaku Smartlab diffractometer with a 9 kW rotating copper anode. All data are recorded in the high-resolution mode employing a collimating parabolic mirror, four-bounce monochromator Ge (220) ($+ - - +$) and a system of collimating slits. The resolution of the scheme $\Delta q_z$ is 0.0004 Å$^{-1}$. All the optical scheme parameters – the beam divergence after the mirror/monochromator/slits optical system, spectral distribution of the incident radiation, and dispersion – are taken into consideration when the experimental data are analyzed. An accurate account of the experimental setup ensures a correct physical description of the system. Otherwise, artificial layers appear in fitting models.[30]

First assessment of the quality of the grown films comes from X-ray diffraction scans. Figure 1 shows $\theta - 2\theta$ diffraction spectrum for Sample A. As far as we know, the only previous reported attempt to grow EuO directly on Si revealing the resulting XRD spectrum is Ref. 22. The remarkable difference is that all peaks in Figure 1 belong to the EuO film or the Si substrate while the XRD scan reported in Ref. 22 shows a significant admixture of bulk europium silicide $EuSi_2$, which is highly detrimental to spin injection. A rough criterion for the quality of the film is the ratio of intensity of a EuO peak to that from the Si substrate. Comparison of Figure 1 with Ref. 22 demonstrates that the former has an order of magnitude larger ratio of intensities of EuO peaks to that of Si (400) than the latter, despite our EuO film being 2 times thinner than that of Ref. 22. Employment of SrO buffer between EuO and Si removes $EuSi_2$ (400) peak in Ref. 22 but the ratio of peak intensities remains inferior to that of Figure 1. In our studies, maximal intensities of EuO and Si reflections correspond to the same orientation of the sample, which excludes a significant angle between lateral atomic planes. Moreover, (202) reflection azimutal $\varphi$-scan demonstrates perfect coincidence of EuO and Si peaks certifying parallel alignment of vertical facets of EuO and substrate.[28]

Before proceeding with X-ray analysis of the grown films it is worth noting that certain structural information can be obtained *in situ*. The growth is controlled with reflection high-energy electron diffractometer fitted with kSA 400 Analytical RHEED System. The lateral lattice parameter can be evaluated by measuring the distances between the intensity maxima of the streaks along the [110] azimuth. Fractional reflexes are not taken into account. The data are calibrated by comparison with the RHEED pattern for clean Si surface with known structure.

Figure 2 shows how initial steps of EuO growth modify the lateral lattice parameter as determined from RHEED images. 10 MLs shown correspond to the low-temperature stage of the growth. Plastic relaxation of strains brings the lattice parameter from that of Si to that of bulk EuO. It is important that RHEED images do not point at any signs of 3D growth.

Motivated by possible spintronic applications, we are predominantly interested in the structure of EuO and its relaxation *at the interface with Si*. The vicinity of the (200) reciprocal lattice point is best suited for the analysis as the EuO (200) peak has the highest intensity. This reflection is allowed for EuO with the rock-salt crystal structure but it is forbidden for Si with the diamond crystal structure. The choice of the (200) peak ensures that our analysis of the epitaxial layer of EuO is not complicated by contributions from silicon.

Figure 3 shows X-ray reflectivity curve and a detailed pattern of X-ray $\theta - 2\theta$ diffraction scan in the vicinity of the EuO (200) peak for Sample A. Well-resolved thickness fringes are observed not only for EuO (200) but also for EuO (400) and EuO (600) peaks. This remarkable result points at sharpness of EuO interfaces and reflects superb structural quality of the film.[31] As far as we know, there are no reports by other groups of thickness fringes in the XRD pattern for EuO thin films.

It is a common practice to perform separate studies of XRR and XRD experimental data though a combined analysis may give a more detailed structure of the interface. The methods have dissimilar sensitivity to structural parameters. When coupled, they provide a set of physical constrains for the structural model, which leads to a better accuracy and unambiguous character of the solution. It is especially beneficial for comprehensive characterization of ordered and disordered layers and interfaces in complex heterostructures. Below, we employ a combined analysis of XRR and XRD data based on the common complex refractive index.

Traditional analysis of XRR curves is based on a layered structural model, its parameters being determined iteratively. Each layer is characterized by its thickness, roughness, and complex refractive index $n$, which is a function of the chemical composition and the material's density. In the limit of small electric susceptibility $\chi$

$$n = 1 - \frac{r_0 \lambda^2}{2\pi} \sum_j n_j f_j, \qquad (1)$$

where $\lambda$ is the wavelength; $r_0$ is the classical electron radius; $n_j$ – the number density and $f_j$ – the atomic scattering factor of the $j$-th atom. This approach does not take into account the crystal structure of the sample. Diffraction from a crystalline structure is commonly described in terms of Fourier components of the electric susceptibility of the crystal:[32]

$$\chi(\mathbf{r}) = \sum_{\mathbf{h}} \chi_{\mathbf{h}} e^{-2\pi i \mathbf{h r}}. \qquad (2)$$

Within the two-wave approximation of the dynamical theory of diffraction, the series is limited by two waves – refracted and diffracted on one group of planes. The former ($\mathbf{h} = \mathbf{0}$) propagates irrespective of the crystallinity of the sample while the latter corresponds to two vectors of the reciprocal lattice $\mathbf{h} = \pm \mathbf{h}_{hkl}$:

$$\chi(\mathbf{r}) = \chi_\mathbf{0} + \chi_\mathbf{h} e^{-2\pi i \mathbf{h r}} + \chi_{\bar{\mathbf{h}}} e^{2\pi i \mathbf{h r}}. \qquad (3)$$

Fourier components $\chi_\mathbf{h}$ are functions of atomic scattering factors:

$$\chi_{\mathbf{h}} = -\frac{r_0 \lambda^2}{\pi V} \sum_j f_j e^{2\pi i (hx+ky+lz)}, \qquad (4)$$

where $V$ is the unit cell volume; x, y, z – fractional coordinates of the $j$-th atom. In particular, Eq. (1) and (4) give simple relation between $n$ and $\chi_\mathbf{0}$:

$$n = 1 + \frac{\chi_\mathbf{0}}{2}. \qquad (5)$$

This linear approximation is valid for small values of $\chi_\mathbf{0}$.

The relation between complex refractive index and electric susceptibility couples XRR and XRD schemes into a single structural model based on calculations of $\chi$. The film is sliced into a sequence of layers characterized by structural parameters: thickness, interlayer distance ($d$), and electric susceptibility ($\chi_\mathbf{0}$, $\chi_\mathbf{h}$). With respect to fitting of an XRR curve, variation of $\chi_\mathbf{0}$ is equivalent to traditional variation of density. In general, an XRR fit is more sensitive to $\chi_\mathbf{0}(z)$ profile than that of XRD as long as the crystalline layer is thin. In contrast, an XRR fit does not depend on $\chi_\mathbf{h}(z)$ profile; it mirrors the Debye-Waller factor thus reflecting disorder in the crystal structure. The built-in coupling of XRR and XRD descriptions comes from common spatial evolution of structural parameters. The regions where parameters change with depth are identified as interfaces. Those regions are modelled by sets of thin layers with thickness close to 2 MLs. The overall thicknesses of interfaces and the bulk of the film are variable parameters of the model.

Analysis of reflectivity curves is based on the Abeles matrix method[33,34] while modelling of XRD rocking curves is carried out in the framework of the dynamical diffraction theory developed for semiconductor multilayer structures.[31] The iterative optimization procedure employs the reduced chi-squared goodness-of-fit statistics. The global optimization is achieved with a differential evolution algorithm successfully adapted to X-ray problems.[35] The result is a self-consistent structural model which describes both XRR and XRD experimental curves presented in Figure 3.

Best-fit depth profiles of interlayer distance $d$ and Fourier components of electric susceptibility for Sample A are shown in Figure 4. Figure 3 demonstrates that the optimal structural model provides a good fit for both XRR and XRD experimental data (the actual value of $\chi^2$ is 1.20). Crystallinity of EuO layers correlates with $\chi_h(z)$ profile. This parameter quite expectedly vanishes for amorphous SiO ($d$ for SiO is well-tabulated in literature). $\chi_h(z)$ is also zero for Si; Si (200) reflection being forbidden. Description of EuO is more informative. The profile is dominated by about 400 Å of EuO with constant values of $d$, $\chi_0$, and $\chi_h$. This region is identified as fully relaxed EuO with the interlayer distance 2.5666(1) Å which corresponds to the bulk value. Arguably, the most important part of Figure 4 is the interface region between relaxed EuO and the Si substrate. It shows gradual relaxation of EuO. The interlayer distance for lamellas of this interface is smaller than that for relaxed EuO because lateral stretching of EuO leads to compression of the vertical lattice parameter to avoid significant changes in the unit cell volume. The value of $d$ (2.33 Å) for the closest to Si lamella is what one would expect taking into account that the first monolayer of EuO is spanned on Si lattice. The thickness of the interface is 27±4 Å, which is about 10 MLs of EuO. It perfectly correlates with *in situ* RHEED image dynamics (see Figure 2).

The other interface (between EuO and SiO) has similar thickness (28±4 Å). The average interlayer distance for this interface is larger than that for relaxed EuO. We associate this with local over-oxidation of a few layers of EuO caused by oxygen diffusion through SiO capping. Indeed, $\chi_0(z)$ profile for SiO gradually decreases starting already at the EuO layer. This behavior is manifested as a low-frequency modulation of Kiessig fringes on the XRR curve. It indicates directly that the capping layer of SiO is uneven and this type of protection of EuO surface from oxidation is insufficient. Thus, lamellas with increased interlayer distance at the SiO/EuO interface effectively model incorporation of local regions of strained $Eu_2O_3$ into the regular lattice of EuO.

Figure 5 shows results of XRR and XRD measurements for Sample B. The diffraction curve has a well-developed peak at $\theta$ close to 15.8°. We associate it with cubic $Eu_2O_3$ used as a capping layer. It also correlates with the RHEED *in situ* observation of coherent oxidation of the EuO surface layer. Thus, the structural model for Sample B is represented by a homogeneous layer of EuO and a layer of $Eu_2O_3$ with parameters depending on the depth. Besides, the model incorporates two interfaces, above and below the EuO layer. Figure 6 presents calculated profiles of $d$, $\chi_0$, and $\chi_h$ for Sample B. Quality of the fit (see Figure 5) is characterized by $\chi^2=1.45$. It is somewhat worse than that for Sample A due to presence of two overlapping peaks. The interfaces between relaxed EuO and Si are quite similar for Samples A and B. It is not surprising because the growth conditions are roughly the same. The other interface (27±5 Å) is marked by changes of the lattice parameter and the structural factor without significant effect on the electron density – layers of EuO and strained $Eu_2O_3$ have similar electron densities and the borderline between them in the $\chi_0(z)$ profile is blurred. The lattice parameter determined from position of $Eu_2O_3$ peak $d_{400}$=2.8285(1) Å ($a$=11.314 Å) is larger than the value for bulk $Eu_2O_3$ by about 4%. It can be explained by tetragonal deformation of the $Eu_2O_3$ layer. On the other hand, such deformation may be caused by a high density of defects:[36,37] the structural factor for $Eu_2O_3$ is significantly smaller than that for EuO indicating low structural quality of the capping layer.

In summary, motivated by perspectives of EuO/Si system for silicon spintronics and recent advances in the epitaxial growth of EuO directly on silicon, we thoroughly studied two EuO films with different protective capping: SiO/EuO/Si and $Eu_2O_3$/EuO/Si. In contrast to previous attempts, we achieve a direct structural coupling across the EuO/Si interface: our XRD $\theta - 2\theta$ scans show absence of unwanted crystalline phases and well-developed thickness fringes for all EuO peaks. The crystalline quality of our films exceeds not only that reported for EuO grown directly on Si but also for EuO grown on a buffer layer of SrO. Both XRR and XRD curves are well fitted with a common structural model. The interface of about 10 MLs of EuO describes relaxation from EuO spanned on Si to the bulk structure in full accordance with *in situ* RHEED data. We determine that amorphous SiO covers EuO film unevenly and its protection of EuO from oxidation is not absolute. Controlled oxidation of EuO leads to defected but crystalline $Eu_2O_3$ capping.

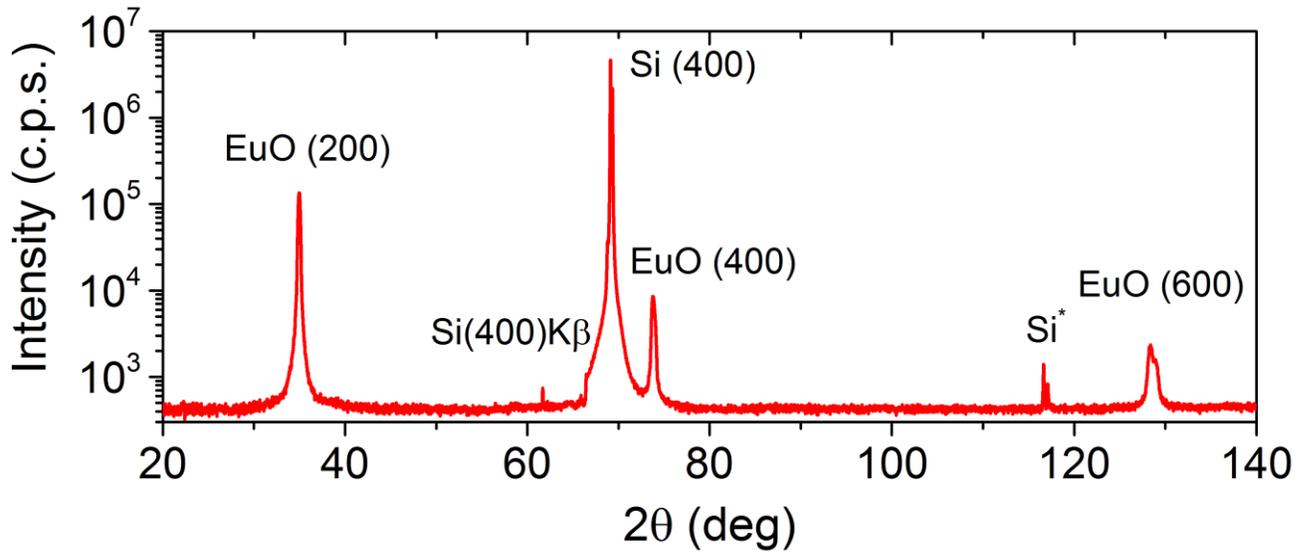

**Figure 1.** X-ray $\theta - 2\theta$ diffraction spectrum of SiO/EuO/Si structure (Sample A) recorded without a monochromator. Si* is a Renninger peak.

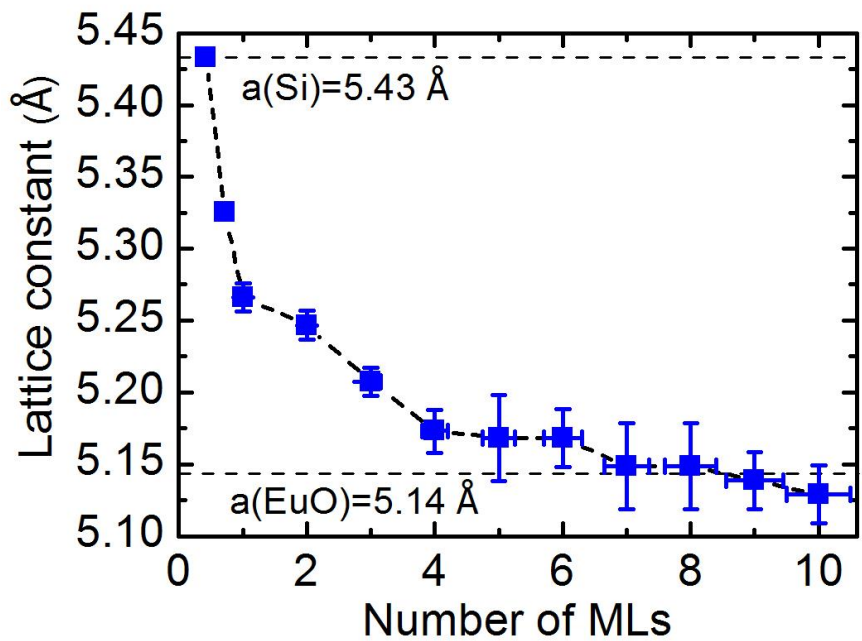

**Figure 2.** Variation of the lateral lattice parameter at the initial stage of the growth as determined by RHEED.

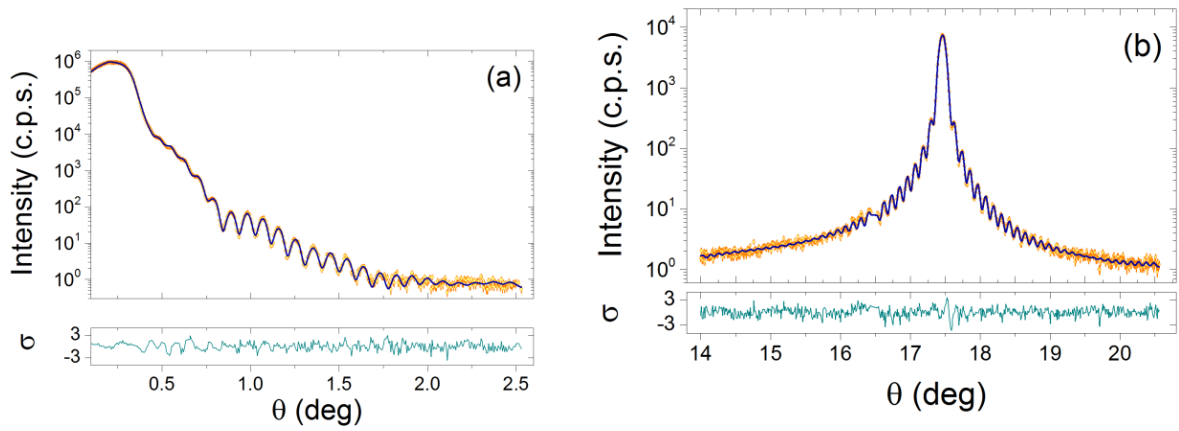

**Figure 3.** Experimental scans (orange) and theoretical fits (blue) for Sample A: a) X-ray reflectivity curve; b) part of X-ray $\theta - 2\theta$ diffraction spectrum near the EuO (200) peak.

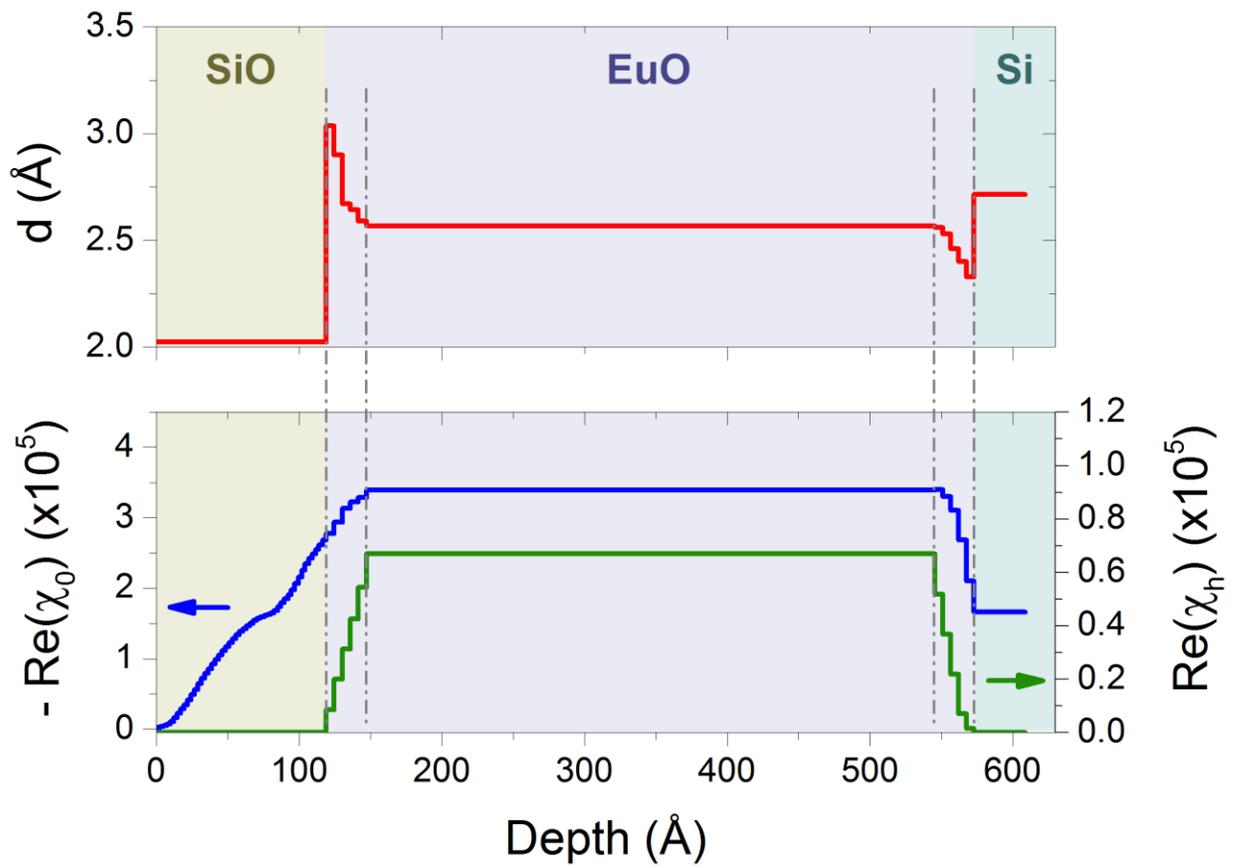

**Figure 4.** Depth profiles of interlayer distance ($d$) and electric susceptibility ($\chi_0$ and $\chi_h$) as determined from combined analysis of XRR and XRD data for sample A.

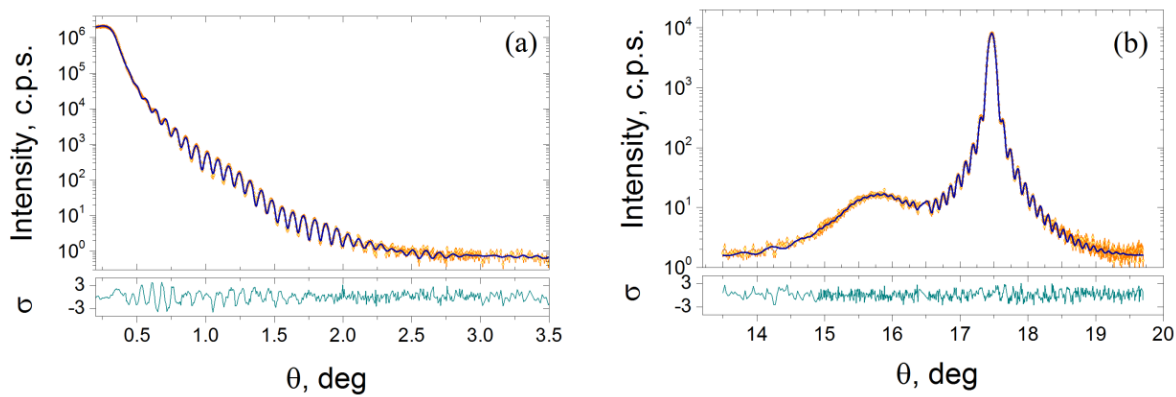

**Figure 5.** Experimental scans (orange) and theoretical fits (blue) for Sample B: a) X-ray reflectivity curve; b) part of X-ray $\theta - 2\theta$ diffraction spectrum near the EuO (200) peak.

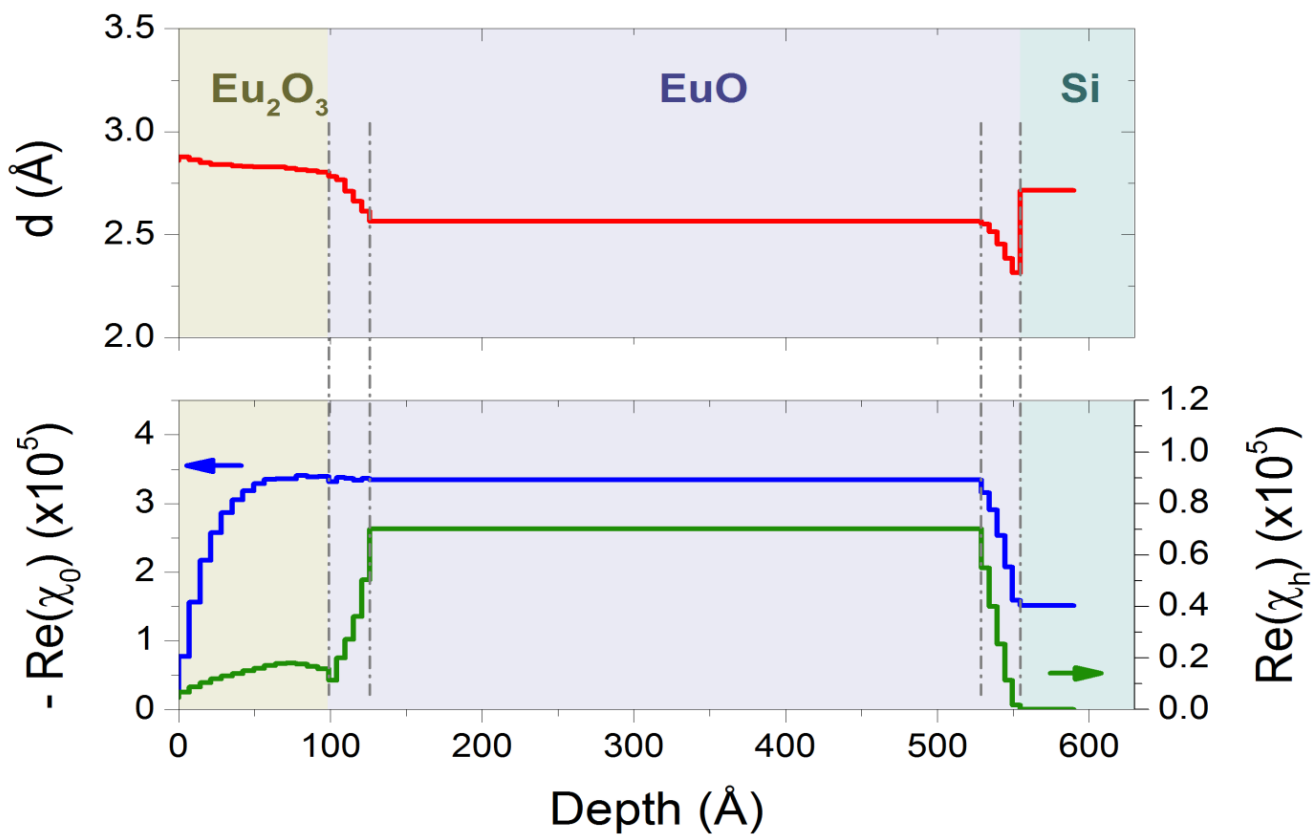

**Figure 6.** Depth profiles of interlayer distance ($d$) and electric susceptibility ($\chi_0$ and $\chi_h$) as determined from combined analysis of XRR and XRD data for sample B.

AUTHOR INFORMATION

**Corresponding Author**

*E-mail: mussr@triumf.ca (VGS).

ACKNOWLEDGMENT

The work is partially supported by NRC "Kurchatov Institute", Russian Foundation for Basic Research through grant 13-07-00095 and Russian Science Foundation through grant 14-19-00662.

For table of contents use only

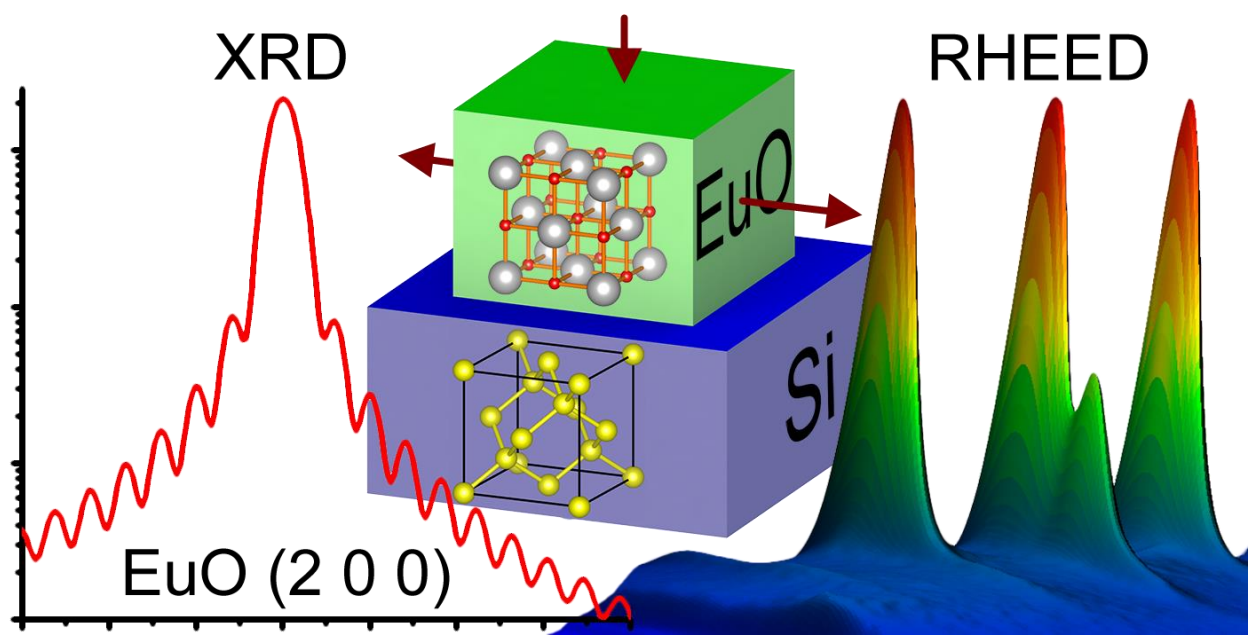